\documentclass[aps,prb,preprint]{revtex4}
\usepackage{amsmath,amssymb,epsfig}

\begin{document}
  

\title{Delocalization effect of the Hubbard repulsion in exact 
terms and two dimensions.}
\author{Zsolt~Gul\'acsi}
\address{Department of Theoretical Physics, University of Debrecen, H-4010 
Debrecen, Hungary }
\date{April 21, 2008}
\begin{abstract}
The genuine physical reasons explaining the delocalization effect
of the Hubbard repulsion $U$ are analyzed. First it is shown that always when 
this effect is observed, $U$ acts on the background of a macroscopic 
degeneracy present in a multiband type of system. After this step I 
demonstrate that acting in such conditions,
by strongly diminishing the double occupancy, $U$ spreads out the 
contributions in the ground state wave function, hence strongly increases 
the one-particle localization length, consequently extends the one-particle 
behavior producing conditions for a delocalization effect. To be valuable,
the reported results are presented in exact terms, being based on the first 
exact ground states deduced at half filling in two dimensions for a prototype 
two band system, the generic periodic Anderson model at finite value of the 
interaction.
\end{abstract}
 \maketitle


\section{Introduction}

The ability of the Hubbard repulsion $U$ to drive transitions to insulating 
behavior is widely studied and known \cite{egy}. However, in the recent years
considerable interest is devoted to the study of the converse phenomenon, 
namely the delocalization effect produced by $U$ \cite{garg,scal,dago}. 
This effect attracts much attention
given by the unexpected nature of the process, since it can finally lead to an 
interaction driven metallic behavior starting from an insulator, instead to 
drive transitions to insulating behavior from a metal, as intuitively is 
expected. The interest for understanding the delocalization effect of $U$
has been first set up by the observed conducting behavior in two dimensional
(2D) disordered systems \cite{dis} where the electrons are localized in the 
starting non-interacting phase by disorder. But, opposed to the random case, 
periodic external potential is also able to cause insulating behavior, hence
can be as well used as a starting point in investigating the delocalization 
effect of $U$. On this line the ionic Hubbard model is 
investigated at half filling in different approximations, 
or numerical methods,
starting from
dynamical mean field theory \cite{garg}, it's cluster version \cite{dago} or 
quantum Monte Carlo technique \cite{scal}, obtaining insulator to metal 
transition in 2D, even if several aspects of the results are under 
debate \cite{dago}. 

On the other side, the clear physical reason for the $U$ produced
delocalization is far to be properly understood. 
Up to this moment only correlation screening of 
the one-body potential responsible for the insulating phase \cite{garg}
has been invoked as an intuitive possibility for explanation.

In order to understand the physical reasons for a possible insulator to metal
transition caused by the Hubbard $U$ we have to pay attention to two aspects
which must concur in producing a conducting behavior starting from an 
insulator \cite{gebhard}: i) the one-particle states must become extended, 
and ii) the gap in the charge excitation spectrum must disappear. The 
correlation screening of the one-body potential can be in principle 
a physical insight 
in explaining point ii), but it does not explain how, given by $U$, point 
$i)$ becomes also satisfied, e.g. how the state becomes extended given by 
the Hubbard interaction in a genuine multielectronic case. The deep
physical reasons of this last process 
are not clearly known, and in my knowledge have not been directly investigated 
so far at a genuine multielectronic level in conditions of interest here,
e.g. around half filling.

The purpose of the presented paper is to provide valuable information 
regarding this issue. During the paper we aim to answer the question: 
Is there a direct evidence that $U$, in the studied circumstances is able to 
extend one-particle states ?  To obtain a valuable answer, the description is 
made here at exact level in 2D. Our strategy is to deduce in a 
non-approximated manner the exact ground states in conditions of interest.
A such obtained state will be a genuine multielectronic state, 
hence usually, is not possible to discuss rigorously about one-particle states
in its frame.  But, based on it, one-particle behavior 
can be tested by calculating the expectation value of
the long-range hopping type of correlations
\begin{eqnarray}
\Gamma({\bf r}) = \frac{1}{N_{\Lambda}} \sum_{{\bf i}=1}^{N_{\Lambda}}
\Gamma_{\bf i}({\bf r}), \quad
\Gamma_{\bf i}({\bf r})=(1/2) \sum_{\sigma}|\langle \hat b^{\dagger}_{{\bf i},
\sigma} \hat b_{{\bf i}+{\bf r},\sigma}+H.c \rangle  |,
\label{Intr1}
\end{eqnarray}
where $N_{\Lambda}$ represents the number of lattice sites, and
$\hat b^{\dagger}_{{\bf i},\sigma}$ creates an electron at site ${\bf i}$
with spin $\sigma$. The study will be given in this paper for the 
non-disordered case in homogeneous conditions, hence $\Gamma({\bf r})
= \Gamma_{\bf i}({\bf r})$ holds. During such a study, the one-particle 
localization length $\ell$ can be also introduced by 
$\Gamma({\bf r}) \sim \exp(- |{\bf r}|/\ell)$. 
In order to answer the question aimed to be investigated here, one simply 
must analyzes whether, or not,
$U$ is able to increase $\Gamma({\bf r})$ and $\ell$ 
in the conditions of interest. 

In order to fix the frame of the study, we must clarify the nature of the
{\it circumstances of interest}. On this line
one observes that in both cases mentioned in the introductory part, 
the $U$ driven
delocalization effect emerges on the background of a macroscopic degeneracy. 
Indeed in the disordered cases the ground states are macroscopically 
degenerate \cite{en1}, while in the ionic Hubbard model case the effects are 
observed in the neighborhood of the point where the single occupancy at 
one-one site of the sublattices A and B have the same energy as an empty site 
in A and a double occupied site in B \cite{scal}. Furthermore, one knows that 
the ionic Hubbard model is of two band type \cite{sorel}, while real 
materials, also disordered once, have a multiband structure which one usually 
projects in a few band picture \cite{volx}. This projection is stopped here 
for convenience at the two band level.  

Given by these observations, one realizes that in order to
find the physical insights of interest here,
one must considers a prototype two band model in 2D. The non-interacting case 
must be set in the position of a macroscopic degeneracy from the point
of view of the one particle states such to find an insulator at $U=0$. 
This can be achieved by choosing the Hamiltonian parameters such to provide a
bare flat (e.g. macroscopically degenerate) lowest band, separated by a finite
gap from the upper dispersive bare band, and fixing the filling of the 
system to half (e.g. lowest band: completely filled, upper band separated by 
a gap: empty). After this step one turns on $U$, and analyzes its effect. 

This experience is quite interesting also
because one knows that  
turning on the interaction in many-body systems with macroscopic degeneracy, 
one always finds a fascinating physics as for example in the case of the 
Laughlin state in two dimensions \cite{intro1}, Hubbard chains with bare 
degenerate bands \cite{intro2}, or geometrically frustrated spin systems 
\cite{intro3}. 

The prototype two band model chosen for the study in this paper will be the 
periodic Anderson model (PAM), simply because several technical aspects 
regarding its ground state construction are known \cite{intro4,intro5}.
The PAM has been constructed to be
the simplest microscopic model used to investigate the characteristic 
properties of heavy fermion and intermediate-valence compounds containing 
elements with incompletely filled f-shells \cite{intro4,intro5}, but serves 
as well, as a test prototype two band model used in various circumstances of 
interest \cite{en1,intro7,intro7a}. The PAM
consists of strongly correlated, almost localized f-electrons which
experience an on-site Coulomb repulsion $U > 0$, and hybridize with a band of 
non-interacting conduction (d-) electrons. In contrast to the one 
dimensional Hubbard model, PAM is not integrable even in dimension 
$D = 1$. Therefore our knowledge of the physics described by the PAM at
finite and nonzero $U$ is almost exclusively obtained by approximate methods. 
Notable exceptions are the constructed exact ground states in dimensions 
$D=1,2,3$, at and below $1/4$, or at and above $3/4$ filling 
\cite{intro8,intro9,intro10,intro4,intro5}, e.g. 
$\bar n =N/(4N_{\Lambda}) \in (0,1/4] \cup [3/4,1)$, where
$N$ is the number of electrons.
The physically quite interesting region situated at and in the vicinity of
half filling, strongly debated at the level of different approximations or
numerical methods
\cite{d1,d2,d3}, lacks exact results for $0<U<\infty$.

The used technique casts the Hamiltonian in a positive semidefinite form 
\cite{intro4}, and the fact that a such expression has the
possible minimum eigenvalue zero, represents the route in obtaining the 
exact ground states \cite{plus}. The method can be applied even in unexpected
situations in the context of exact solutions, as \cite{intro5} $D=3$,
disordered and interacting systems in 2D \cite{en1}, non-integrable
Hubbard chains in external fields \cite{intro2}, or stripe, 
checkerboard and droplet ground states \cite{intro7} in 2D. However, given
by strong correlation effects not treatable mathematically up to this moment
in the studied concentration region, it has not yet been applied for the 
PAM at finite $U \ne 0$ around half filling. 

The deduced results provide the first exact ground states for the 2D generic
PAM at half filling and finite value of the interaction. This is achieved by
the use of extended operators in constructing the ground state, technique 
which is described also for the first time in the context of 
non-integrable systems in this paper. 

The comparison of ground states at $U=0$ and $U > 0$ on the line of 
(\ref{Intr1})
shows that the Hubbard repulsion, by strongly diminishing the double occupancy,
spread out the contributions in the ground state, hence is able to 
increase by several order of magnitudes the one-particle localization length.
Hence the one-particle behavior in the analyzed conditions is extended by $U$,
which is a clear physical source of a delocalization effect. Consequently, the
answer to the question we are looking for in this paper is: yes. 
I also note that one already 
knows that even away from half-filling, $U$ is able to create in relatively
similar conditions subtle 
correlated conducting phases \cite{intro2,intro4,intro5}.
It is also known that insulator to metal transition can be driven as well by
disorder \cite{intro6}, the process emerging at half filling again from a
strongly degenerate insulating phase. 

The remaining part of the paper is structured as follows. Section II. 
describes the model and its solution in the non-interacting case, Section III.
prepares the deduction by transforming the starting Hamiltonian in a positive 
semidefinite form, Section IV. describes the extended operators that allow the
construction of the exact ground states, Section V. presents the deduced
exact ground state, characterizing as well the physical properties of the 
system, while Section VI. containing the summary and conclusions closes the 
presentation.  

\section{The model used}

\subsection{The Hamiltonian}

I analyze a generic PAM Hamiltonian $\hat H= \hat H_0 + U \hat U_f$
defined on an $N_{\Lambda}=L \times L$, 2D Bravais lattice 
with primitive vectors $({\bf x},{\bf y})$. The Hubbard term is 
$\hat U_f = \sum_{\bf i} \hat n^f_{{\bf i},\uparrow} \hat n^f_{{\bf i},
\downarrow}$, arbitrary $U > 0$ is considered, and
\begin{eqnarray}
&&\hat H_0 = \sum_{{\bf i},\sigma} \Big\{ (\sum_{{\bf r}_d} t_{{\bf r}_d}
\hat d^{\dagger}_{{\bf i},\sigma} \hat d_{{\bf i}+{\bf r}_d,\sigma} 
+ V_0 \hat d^{\dagger}_{{\bf i},\sigma} \hat f_{{\bf i},\sigma} + 
\label{eq1}
\\
&&\sum_{{\bf r}={\bf x},{\bf y}} \sum_{b,b'=d,f; b \ne b'}
V_{\bf r}^{b,b'} \hat b^{\dagger}_{{\bf i},\sigma} \hat b'_{{\bf i}+{\bf r},
\sigma} + H.c.) + E_f \hat f^{\dagger}_{{\bf i},\sigma} \hat f_{{\bf i},\sigma}
\Big\},
\nonumber
\end{eqnarray}
where in order to describe real systems, the hopping matrix element $t_{
{\bf r}_d}$ covers by the index ${\bf r}_d$ nearest and 
next nearest neighbors, $V_0$
and $V^{b,b'}_{\bf r}$, $b,b'=d,f$, give the strengths of the on-site and 
nearest neighbor hybridizations, and $E_f$ represents the on-site 
$f$-electron energy. 
During calculations periodic boundary conditions are used, and the filling
is fixed to half (e.g. $\bar n = 1/2 $, $N=2N_{\Lambda}$).

\subsection{The $U=0$ non-interacting case}

In order to have a macroscopic degeneracy at one-particle level, one considers
first the non-interacting $U = 0$ case, and selects the $\hat H_0$ parameters 
such to obtain a lowest flat band in the non-interacting band structure.
For simplicity one considers the $(x,y)$ symmetric case $t_1=t_{\bf x}= 
t_{\bf y}$, $t_2/2=t_{2 {\bf x}}=t_{2 {\bf y}}=t_{{\bf y}\pm {\bf x}}/2$,
$V_1=V^{b,b'}_{\bf x}=V^{b,b'}_{\bf y}$. To have a lowest flat band in the 
non-interacting band structure one takes
\begin{eqnarray}
V_1 t_1 = V_0 t_2, \quad E_f = 2 \frac{V_1^2}{t_2} -\frac{V_0^2 t_2}{2 V_1^2}
- 2 t_2, \quad t_2 > 0.
\label{eq2}
\end{eqnarray}
Indeed, for conditions (\ref{eq2}), $\hat H_0$ from (\ref{eq1}) transformed in
${\bf k}$ space, can be diagonalized into the form
\begin{eqnarray}
\hat H_0 = \sum_{\sigma} \sum_{\bf k}[E_{1,{\bf k}} \hat C^{\dagger}_{1,
{\bf k},\sigma} \hat C_{1,{\bf k},\sigma} + E_{2,{\bf k}} 
\hat C^{\dagger}_{2,{\bf k},\sigma} \hat C_{2,{\bf k},\sigma}],
\label{eq3}
\end{eqnarray}
where $\hat C^{\dagger}_{\nu,{\bf k},\sigma}$ represents the canonical Fermi 
operator of the $\nu$th diagonalized band (here $\nu=1,2$) of $\hat H_0$.
In the case of (\ref{eq2}), the two bands $E_{1,{\bf k}}=-(t_2/2)[(V_0^2/V_1^2)
+4] < E_{2,{\bf k}}=E_{1,{\bf k}} + 2t_2[ (V_1/t_2)^2+(C_{\bf k} +
V_0/(2V_1))^2]$, $C_{\bf k}= \cos ({\bf x}{\bf k}) + \cos ({\bf y}{\bf k})$,
are clearly separated, the lowest band is flat, the band gap being 
$\Delta = 2 t_2 [(V_0/t_1)^2 +(|V_0/2V_1|-2)^2] \ne 0$. Consequently, in the
non-interacting case, given by the presence of the lowest bare flat band, a
macroscopic degeneracy is present, whose degree at the one particle level is
$2 N_{\Lambda}$.

The canonical Fermi operators present in (\ref{eq3}) are given by
\begin{eqnarray}
&&\hat C^{\dagger}_{1,{\bf k},\sigma}=\frac{1}{\sqrt{2}} [ \sqrt{1+F_{\bf k}} 
\: \: \hat d^{\dagger}_{{\bf k},\sigma} - sign(\alpha_{\bf k})
\sqrt{1-F_{\bf k}} \: \: \hat f^{\dagger}_{{\bf k},\sigma}],
\nonumber\\
&&\hat C^{\dagger}_{2,{\bf k},\sigma}=\frac{1}{\sqrt{2}} [ \sqrt{1-F_{\bf k}} 
\: \: \hat d^{\dagger}_{{\bf k},\sigma} + sign(\alpha_{\bf k})
\sqrt{1+F_{\bf k}} \: \: \hat f^{\dagger}_{{\bf k},\sigma}],
\label{eq4}
\end{eqnarray} 
where $\alpha_{\bf k}= V_0+2V_1 C_{\bf k}$, 
$\beta_{\bf k} = V_1^2/t_2-t_1^2/(4t_2)-t_1 C_{\bf k} -t_2 C^2_{\bf k}$,
and $F_{\bf k}=[1+(\alpha_{\bf k}/\beta_{\bf k})^2]^{-0.5}$.

The ground state at $U=0$  and
half filling (lowest band completely filled, upper band completely empty) 
becomes 
\begin{eqnarray}
|\Psi^0_g\rangle = \prod_{\sigma}
\prod_{{\bf k}=1}^{N_{\Lambda}} \hat C^{\dagger}_{1,{\bf k},\sigma} |0\rangle,
\label{eq5}
\end{eqnarray}
which represents a paramagnetic band insulator. The ground state energy
corresponding to (\ref{eq5}) is $E_g^0(2L^2)=- t_2 L^2[(V_0^2/V_1^2)+4]$.

Denoting by $\langle ...\rangle_0$ ground state 
expectation values calculated in terms of (\ref{eq5}), for each site 
${\bf i}$ of the system one finds $d^f_{\bf i}=\langle \hat n^f_{{\bf i},
\uparrow} \hat n^f_{{\bf i},\downarrow} \rangle_0 \ne 0$. Concerning 
$\Gamma({\bf r})$ from (\ref{Intr1}), using (\ref{eq5}),
$b=d,f$, one finds $\Gamma({\bf r}) \sim \exp(-|{\bf r}|/\ell)$,
where the associated one particle localization length $\ell$ is of order unity
in lattice constant units (for example, for ${\bf r} \parallel {\bf x}$, 
one obtains $\ell/|{\bf x}|=0.5$ for the Hamiltonian parameters fixed in 
Fig.8.). Based on this result one concludes that the non-interacting
system in the studied conditions represents a localized band insulator.
 
\section{The Hamiltonian transformed in a positive semidefinite form}

\subsection{The transformed Hamiltonian}

In order to deduce the exact ground state in the interacting case, one first
transforms the Hamiltonian $\hat H =\hat H_0 + U \hat U_f$, $U > 0$, 
given in (\ref{eq1},\ref{eq2}) in a positive semidefinite form. I note that 
this exact transformation is $N$ independent, and provides
\begin{eqnarray}
&&\hat H = \hat H_0 + U \hat U_f = \hat G + U \hat U_f +E_g(N),
\nonumber\\
&&\hat G =\sum_{v=1,2} \sum_{e=\pm 1} \sum_{{\bf i} \in S_v} 
\hat A^{\dagger}_{{\bf i},v,e} \hat A_{{\bf i},v,e}, 
\quad E_g(N)=-K_d N.
\label{eq6}
\end{eqnarray}
To clarify Eq.(\ref{eq6}), the following explications must be given:
In the first line of (\ref{eq6}), $\hat G$ and $U \hat U_f$ are positive
semidefinite operators, while $E_g(N)$ represents the ground state 
energy (see Sec.IV.), a number which has a given dependence on the Hamiltonian 
parameters, and whose expression will be clarified below. In order to obtain 
(\ref{eq6}), first we divide the lattice in two sublattices.
The sublattices are denoted by $v=1,2$, and all lattice sites from the
sublattice $v$ are denoted by $S_v$. After this step one introduces the cell
operators $\hat A_{{\bf i},v,e}$, which are given by
\begin{eqnarray}
\hat A_{{\bf i},v,e} = a_{1,f} \hat F_{{\bf i},\lambda_v,e} 
+ \sum_{n=1}^5a_{n,d} \hat D_{{\bf i}+{\bf r}_n,\lambda_v,e}.
\label{eq7}
\end{eqnarray}
Even if labelled by the site ${\bf i}$, the $\hat A_{{\bf i},v,e}$ operators
are not local, and collect contributions in the form of a linear
combination, from five sites situated around the site ${\bf i}$. These five
sites denoted by ${\bf i}+{\bf r}_n$, $n=1,2,3,4,5$, build up together the 
cell defined at the site ${\bf i}$, where taken for increasing $n$, 
${\bf r}_n$ has the values $0,-{\bf y}, {\bf x}, {\bf y}, -{\bf x}$
(see Fig.1).
\begin{figure}[h]
\epsfxsize=6cm
\centerline{\epsfbox{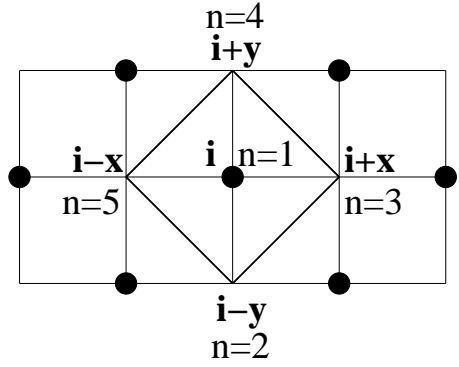}}
\caption{The cell defined at site ${\bf i}$ (thick line) 
containing 5 sites of the 
lattice (${\bf i}, {\bf i}\pm{\bf x},{\bf i}\pm{\bf y}$) on which the 
operator $\hat A_{{\bf i},v,e}$ is defined.
The sites of the sublattice $v$, ${\bf i} \in S_v$,
are denoted by black dots, $({\bf x},{\bf y})$ are the primitive vectors
of the lattice, and $n=1,2,...,5$ represents the notation of sites inside
the cell. }
\label{fig1}
\end{figure} 

After this step one presents below separately in details the 
numerical prefactors, and the operators present in (\ref{eq7})

\subsection{The numerical prefactors of $\hat A_{{\bf i},v,e}$}

The numerical coefficients $a_{1,f}, a_{n,d}$ in (\ref{eq7}) are related
to the Hamiltonian parameters of $\hat H_0$, being given by the 
following system of equations
\begin{eqnarray}
&&t_{\bf x} = a^{*}_{1,d}a_{3,d} + a^{*}_{5,d}a_{1,d},
\quad
t_{\bf y} = a^{*}_{1,d}a_{4,d} + a^{*}_{2,d}a_{1,d},
\nonumber\\
&&t_{{\bf y}\pm {\bf x}} = a^{*}_{2,d}a_{4 \mp 1,d} + a^{*}_{4 \pm 1,d}a_{4,d},
\quad
V_0 = a^{*}_{1,d}a_{1,f}, 
\label{eq8}
\\
&&t_{2{\bf x}} = a^{*}_{5,d}a_{3,d}, \quad
t_{2{\bf y}} = a^{*}_{2,d}a_{4,d}, \quad
V^{df}_{{\bf x}} = a^{*}_{5,d}a_{1,f}, 
\nonumber\\
&&V^{df}_{{\bf y}} = a^{*}_{2,d}a_{1,f}, \quad
V^{fd}_{{\bf x}} = a^{*}_{1,f}a_{3,d}, \quad
V^{fd}_{{\bf y}} = a^{*}_{1,f}a_{4,d}, 
\nonumber\\
&&E_f=|a_{1,f}|^2 - K_d,  \quad K_d=\sum_{n=1}^5 |a_{n,d}|^2.
\nonumber
\end{eqnarray}
It is important
to note that the exact transformation presented in (\ref{eq6}) works only if
the ,,matching conditions'' (\ref{eq8}) are satisfied. I also mention that
the system of equations (\ref{eq8}), through the expression of $K_d$, fixes
as well the value of the ground state energy $E_g(N)$ in (\ref{eq6}).   
One obtains the system of equations (\ref{eq8}) by 1) calculating  
$\hat A^{\dagger}_{{\bf i},v,e} \hat A_{{\bf i},v,e}$ based on 
(\ref{eq7},\ref{eq10}), 2) deducing
$\hat G + E_g(N)$ in (\ref{eq6}) starting from the previous result,
and 3) finally equating to $\hat H_0$ from (\ref{eq1}) the obtained expression.
I note that the presence of $\hat f_{{\bf j},\sigma}$ operator only in
the middle of the cell operator $\hat A_{{\bf i},v,e}$ in
(\ref{eq6}) (see also Fig.1), leads to the absence of the direct $f$-electron
hopping in the transformed Hamiltonian.

The matching conditions (\ref{eq8}), in the $(x,y)$ symmetric case specified
above Eq.(\ref{eq2}), give solution only when (\ref{eq2}) is satisfied.
The solution is characterized by $a_{n \geq 2,d}=a_{2,d}$, and one obtains
\begin{eqnarray}
a_{1,f} = V_1 \frac{\sqrt{2}}{\sqrt{t_2}} e^{i\phi}, \quad
a_{1,d} = \frac{V_0}{V_1} \frac{\sqrt{t_2}}{\sqrt{2}} e^{i\phi}, \quad
a_{2,d} = \frac{\sqrt{t_2}}{\sqrt{2}}  e^{i\phi},
\label{eq9}
\end{eqnarray}
where $\phi$ is an arbitrary phase.

\subsection{The operators contained in $\hat A_{{\bf i},v,e}$}

The operators $\hat D_{{\bf j},\lambda_v,e}$, $\hat F_{{\bf j},\lambda_v,e}$,
$e=\pm 1$, are new genuine canonical Fermi operators substituting the starting
canonical Fermi operators $\hat d_{{\bf j},\sigma}$, $\hat f_{{\bf j},\sigma}$,
$\sigma=\uparrow,\downarrow$. Their expression is given by the following 
relations
\begin{eqnarray}
\hat B_{{\bf j},\lambda_v,+1}= \frac{|\lambda_v|}{
\sqrt{1+|\lambda_v|^2}} (\hat b_{{\bf j},\uparrow} + \frac{1}{
\lambda_v} \hat b_{{\bf j},\downarrow}), \quad
\hat B_{{\bf j},\lambda_v,-1}=\frac{|\lambda_v|}{
\sqrt{1+|\lambda_v|^2}}(\frac{1}{\lambda^{*}_v} \hat b_{{\bf j},\uparrow}
-\hat b_{{\bf j},\downarrow}),
\label{eq10}
\end{eqnarray}
where for $b=d$ ($f$), one has $B=D$ ($F$),
and $\lambda_v$ is an arbitrary complex number.

Looking for the physical meaning of $\hat B$ operators, 
taking for example
the $b=d$ case, using the fermionic spin representation for $1/2$ spin 
operators $\hat S^z_{d,{\bf i}} =(1/2)(\hat n^d_{{\bf i},\uparrow} -
\hat n^d_{{\bf i},\downarrow}), \hat S^{+}_{d,{\bf i}}= \hat d^{\dagger}_{
{\bf i},\uparrow} \hat d_{{\bf i},\downarrow},  \hat S^{-}_{d,{\bf i}}= 
\hat d^{\dagger}_{{\bf i},\downarrow} \hat d_{{\bf i},\uparrow},
\hat S^{\pm}_{d,{\bf i}}=\hat S^x_{d,{\bf i}} \pm i \hat S^{y}_{d,{\bf i}}$,
${\hat S}^2_{d,{\bf i}}=\hat S^z_{d,{\bf i}}\hat S^z_{d,{\bf i}} +(1/2)(
\hat S^{+}_{d,{\bf i}}\hat S^{-}_{d,{\bf i}}+\hat S^{-}_{d,{\bf i}}
\hat S^{+}_{d,{\bf i}})$,
and introducing the states $|D_{{\bf i},\lambda,e}\rangle = \hat D^{\dagger}_{
{\bf i},\lambda,e} |0\rangle$, $\langle D_{{\bf i},\lambda,e} | D_{{\bf i},
\lambda,e}\rangle =1$, one finds for arbitrary $\lambda$
\begin{eqnarray}
{\hat S}^2_{d,{\bf i}}|D_{{\bf i},\lambda,e}\rangle = \frac{3}{4}
|D_{{\bf i},\lambda,e}\rangle, \quad
(\hat  n^d_{{\bf i},\uparrow} + \hat n^d_{{\bf i},\downarrow}) |D_{{\bf i},
\lambda,e}\rangle = |D_{{\bf i},\lambda,e}\rangle.
\label{eq11}
\end{eqnarray}
Since similar relations can also be written for the $b=f$ case,
one concludes that
$\hat B^{\dagger}_{{\bf i},\lambda,e}$ introduces one 
spin $1/2$ fermion ($b=d$, or $b=f$ for $B=D$ or $B=F$) on the site ${\bf i}$.
One further observes that at the level of expectation 
values, the spin direction and orientation are provided by $\lambda$, and $e$. 
Indeed,
for all $\alpha=x,y,z$ one has $\langle D_{{\bf i},\lambda,-1} |
\hat S^{\alpha}_{d,{\bf i}} | D_{{\bf i},\lambda,-1}\rangle = -
\langle D_{{\bf i},\lambda,1} |\hat S^{\alpha}_{d,{\bf i}} | D_{{\bf i},
\lambda,1}\rangle$, and
\begin{eqnarray}
\langle D_{{\bf i},\lambda,1} |\hat S^{\alpha}_{d,{\bf i}} | D_{{\bf i},
\lambda,1}\rangle =\frac{1}{1+|\lambda|^2} [ \delta_{x,\alpha} Re(\lambda) +
\delta_{y,\alpha} Im(\lambda) + \delta_{z,\alpha} \frac{|\lambda|^2-1}{2} ],
\label{eq12}
\end{eqnarray}
such that $(\langle D_{{\bf i},\lambda,e} |\hat {\vec S}_{d,{\bf i}} | 
D_{{\bf i},\lambda,e}\rangle)^2=1/4$.
These results show that in fact $\lambda$ fixes an arbitrary axis along 
which the spin is placed, while $e=\pm 1$ gives the spin direction along 
the fixed 
axis. Indeed, for example at $\lambda=1$ one has
$\hat S^{x}_{d,{\bf i}}| D_{{\bf i},\lambda=1,e}\rangle=(e/2) |D_{{\bf i},
\lambda=1,e}\rangle$, for $\lambda=i$ one obtains
$\hat S^{y}_{d,{\bf i}}| D_{{\bf i},\lambda=i,e}\rangle=(e/2) |D_{{\bf i},
\lambda=i,e}\rangle$, while $\lambda \to \infty$ corresponds to the
$\hat S^{z}_{d,{\bf i}}| D_{{\bf i},\lambda,e}\rangle=(e/2) |D_{{\bf i},
\lambda,e}\rangle$ case (e.g. for $\lambda =1$, $\lambda=i$, and 
$\lambda \to \infty$, the spin axis becomes in order the x,y, and z axis).

I further note that the relation $\hat D^{\dagger}_{{\bf i},\lambda,1} 
\hat D^{\dagger}_{{\bf i},\lambda,-1}= \hat d^{\dagger}_{{\bf i},
\downarrow} \hat d^{\dagger}_{{\bf i},\uparrow}$ holds for all $\lambda$.
It is also important to mention that in Eq.(\ref{eq6}), after the 
transformation of the starting Hamiltonian defined by (\ref{eq1}), for 
$v=1,2$, one obtains arbitrary, but fixed $\lambda_1$ and $\lambda_2$.

\section{Description of the mathematical solution that leads to the
ground state wave function at half filling and $U > 0$.}

In (\ref{eq6}) one has the Hamiltonian transformed in a positive semidefinite
form which now must be analysed in order to deduce its ground state at half
filling in the interacting $U > 0$ case. As mentioned above (\ref{eq1}),
$\hat H$ has been defined on an $L \times L$ Bravais lattice, and obtaining the
transformed form in (\ref{eq6}), the lattice has been divided in two 
sublattices $v=1,2$ containing each $L^2/2$ sites. 

In deducing the ground state at $N$ particles we have to find
a wave vector $|\phi(N)\rangle = \hat K^{\dagger}|0\rangle$, where 
$\hat K^{\dagger}$ introduces $N$ electrons into the system, and based on 
(\ref{eq6})
\begin{eqnarray}
\hat G |\phi(N)\rangle = \hat U_f |\phi(N)\rangle =0 
\label{eq13}
\end{eqnarray}
holds. In such conditions $|\phi(N)\rangle$ represents the ground state of the
$N$-electron system described by $\hat H$ in (\ref{eq6}), the ground state 
energy being $E_g(N)$. In order to prove (\ref{eq13}), one 
shows $\{\hat A_{{\bf i},v,e}, \hat K^{\dagger} \}=0$ for the first part,
and demonstrate that $|\phi(N)\rangle$ not contains double $f$-occupancy
for the second. 

Below I describe the $\hat K^{\dagger}$ operator which will provide the
ground state at $N=2L^2$ (half filling). Since in this concentration region 
strong correlation effects are present, $\hat K^{\dagger}$ has a relatively 
complicated form. Because of this fact, in describing $\hat K^{\dagger}$
first I present it, describe it in details, give its expression, and
explain how it was deduced.

\subsection{The mathematical key of the solution}

Taking $|0\rangle$ as
the bare vacuum, one considers the state 
\begin{eqnarray}
|\Psi^{\lambda_1,\lambda_2}_{e_1,e_2} \rangle = 
\prod_{v=1}^2 \prod_{{\bf i}\in 
S_v} (\hat W^{\dagger}_{v,{\bf i},\tau=1,\lambda_v, e_v}
\hat W^{\dagger}_{v,{\bf i},\tau=4,\lambda_v,e_v}) |0\rangle ,
\label{eq14}
\end{eqnarray}
where 
$\hat W^{\dagger}_{v,{\bf i},\tau,\lambda_v,e_{v}}$ is the broken
line operator defined for the sublattice $v$, labelled by the site ${\bf i}$ 
of $S_v$, breaking direction at the site ${\bf i}$ $\tau_{\bf i}=\tau$, 
spin direction given by $\lambda_v$, and spin orientation specified by 
$e_{v}$. 

One arrives to the broken line operator notion by  looking for the desired 
$2L^2$ number of linearly independent operators suitable in constructing
the ground state at $\bar n=1/2$. If we would like to use 
an operator of the type $\hat C^{\dagger}_{1,{\bf k},\sigma}$ 
(or its Fourier transform in ${\bf r}$ space) from
(\ref{eq4},\ref{eq5}) for this job, we shortly 
realize that only $L^2$ such operators are available in constructing 
the ground  state at $U > 0$, namely those with fixed $\sigma$, since 
mixing the spin index, automatically double occupancy occurs which raises the
energy. Hence, the expression of the interacting ground state requires
new operators $\hat W^{\dagger}$ which must be find. The search for
these leads first to operators constructed along lines in ${\bf y}+{\bf x}$ 
direction. These resemble to the operators presented in Fig.7, but their
number is of order $L$, hence is small. At this point one remembers that by 
breaking a straight line, a site (e.g. the breaking point) becomes attached to
the line, the number of sites being $L^2$. Hence in this manner is possible
to reach the number of desired operators. The presence of the cusp point
on the line complicates a little the expression of the operators, but this 
is in fact the way one reaches the broken line operator notion.
 
Since the structure of the $W^{\dagger}$ operator is not transparent 
in (\ref{eq14}), I explain it in details below.

\subsection{Straight lines and broken lines}

Connecting the sites of a given sublattice in ${\bf y}+{\bf x}$ 
direction,
$n \in [1,L/2]$ different straight lines $l_{{\bf y}+{\bf x},v,n}$ can be 
drawn, each containing $L$ sites, see Fig.2. For simplicity one considers below
even $L$.
\begin{figure}[h]
\epsfxsize=6cm
\centerline{\epsfbox{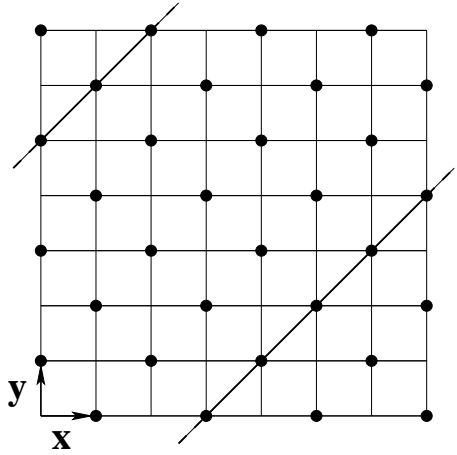}}
\caption{An $8\times 8$ lattice (e.g. $L=8$ holds), containing one straight 
line $l_{{\bf y}+{\bf x},v=1,n}$, (thick line). The direction of the line is
${\bf x}+{\bf y}$, it is contained in the sublattice $v=1$ denoted by 
full circles, the line contains $L=8$ sites, and note that periodic 
boundary conditions are considered.
(${\bf x}$,${\bf y}$) are the primitive vectors of the 2D Bravais lattice.
In the ${\bf x}+{\bf y}$ direction $L/2=4$ such lines can be plotted. }
\label{fig2}
\end{figure}

Now one breaks the straight lines at an arbitrary 
site ${\bf i} \in l_{{\bf y}+{\bf x},v,n}$. In order to be this possible,
one first defines 4 breaking 
directions denoted by the index $\tau$ as shown in Fig.3.
\begin{figure}[h]
\epsfxsize=6cm
\centerline{\epsfbox{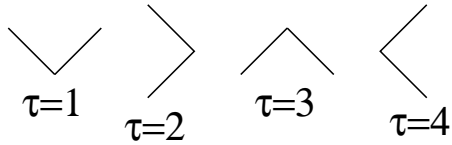}}
\caption{The four possible breaking directions denoted by the
index $\tau$.}
\label{fig3}
\end{figure}
If one fixes the breaking direction at ${\bf i}$ by $\tau_{\bf i}=1$, one
finds the ,,broken line'' as presented in Fig.4. 
\begin{figure}[h]
\epsfxsize=6cm
\centerline{\epsfbox{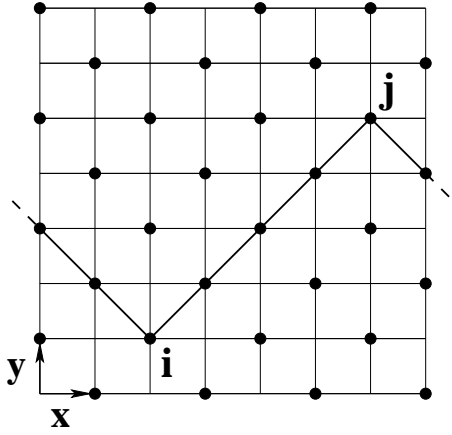}}
\caption{A broken line $l_v({\bf i},\tau_{\bf i}=1)$ (thick line)  defined 
at the site ${\bf i}$ in a $8\times 8$ lattice. 
$l_v({\bf i},\tau_{\bf i}=1)$
connects sites of the $v$ sublattice (full dots). 
Note that $L^2/2$ such broken lines can be plotted for a fixed $v$.}
\label{fig4}
\end{figure}
This is simply obtained by taking the straight line $l_{{\bf y}+{\bf x},v=1,n}$
crossing the site ${\bf i}$, and breaking the line at the site ${\bf i}$ in
the breaking direction $\tau_{\bf i}=1$.
Given by the periodic boundary conditions, a broken line besides its first 
cusp at ${\bf i}$, and $\tau_{\bf i}=1$ (obtained by breaking the starting
straight line at ${\bf i}$), will have a second cusp point 
at ${\bf j}={\bf i}+(L/2)({\bf x}+{\bf y})$, and $\tau_{\bf j}
=3$ (see Fig.4), which is unambiguously determined by 
$({\bf i}, \tau_{\bf i})$. Hence the obtained broken lines can be denoted by
$l_v({\bf i},\tau_{\bf i}=1)$, and their number for fixed $v$ becomes $L^2/2$. 
Consequently, the broken line $l_v({\bf i},\tau_{\bf i}=1)$ is a continuous 
line in the sublattice $v$, containing $L$ sites from $S_v$ and two cusps as 
presented in Fig.4, the cusp characterized by the breaking direction $\tau=1$ 
being present at the site ${\bf i}$.

\subsection{The $W^{\dagger}$ operators}

The $\hat W^{\dagger}_{v,{\bf i},\tau,\lambda_v,e_{v}}$ operator 
represents a linear combination of operators acting on 
${\bf i}'$ sites placed in an extended region surrounding the broken line
$l_v({\bf i},\tau_{\bf i}=\tau)$ as presented in Figs.5-7 for $\tau=1$. In
these figures, the black, open, and gray circles plotted at a 
site ${\bf i}'$ are representing in order
$\hat P^{\dagger}_{{\bf i}',\lambda_v,
e_{v}}= x^d_{{\bf i}'} \hat D^{\dagger}_{{\bf i}',\lambda_v,e_{v}} + 
x^f_{{\bf i}'} \hat F^{\dagger}_{{\bf i}',\lambda_v,e_{v}}$, $\hat D^{
\dagger}_{{\bf i}',\lambda_v,e_{v}}$, and $\hat F^{\dagger}_{{\bf i}',
\lambda_v,e_{v}}$ operators.

The coefficients present in Figs.5-7 are numerical prefactors. 
In order to obtain the mathematical expression of the
$\hat W^{\dagger}_{v,{\bf i},\tau=1,\lambda_v,e_{v}}$ operator, 
one simply adds all contributions from Figs.5-7 as follows
\begin{eqnarray}
\hat W^{\dagger}_{v,{\bf i},\tau =1,\lambda_v,e_{v}} &=& \xi \hat F^{\dagger}_{
{\bf i}-2{\bf y},\lambda_v,e_{v}} +y_0 \hat D^{\dagger}_{{\bf i}-
{\bf y},\lambda_v,e_{v}} +\epsilon \hat D^{\dagger}_{{\bf i}+{\bf y},
\lambda_v,e_{v}} + \mu \hat F^{\dagger}_{{\bf i}+2{\bf y},\lambda_v,e_{v}} 
\nonumber\\
&+& \sum_{n=0}^M (x_n^d \hat D^{\dagger}_{{\bf i}+n({\bf x}+
{\bf y}),\lambda_v,e_{v}} + x_n^f \hat F^{\dagger}_{{\bf i}+n({\bf x}+
{\bf y}),\lambda_v,e_{v}}) 
\nonumber\\
&+&\sum_{n=1}^{M-1} (x_n^d \hat D^{\dagger}_{{\bf i}+n({\bf y}-
{\bf x}),\lambda_v,e_{v}} + x_n^f \hat F^{\dagger}_{{\bf i}+n({\bf x}+
{\bf y}),\lambda_v,e_{v}}) 
\nonumber\\
&+&\sum_{n=1}^{M-1} y_n \hat D^{\dagger}_{{\bf i}+n{\bf x}+(n-1){\bf y},
\lambda_v,e_{v}} + \sum_{n=1}^{M-1} y_n \hat D^{\dagger}_{{\bf i}-
n{\bf x}+(n-1){\bf y},\lambda_v,e_{v}}
\nonumber\\
&+& \sum_{n=2}^{M} y_n \hat D^{\dagger}_{{\bf i}+(n-1){\bf x} + n{\bf y},
\lambda_v,e_{v}} + \sum_{n=2}^{M} y_n \hat D^{\dagger}_{{\bf i}-(n-1)
{\bf x} + n{\bf y},\lambda_v,e_{v}}
\nonumber\\
&+&\gamma \hat D^{\dagger}_{{\bf j}-{\bf y},\lambda_v,e_{v}}
+\beta \hat D^{\dagger}_{{\bf j}+{\bf y},\lambda_v,e_{v}} +
\delta \hat F^{\dagger}_{{\bf j}-2{\bf y},\lambda_v,e_{v}} 
\nonumber\\
&+&
\alpha_1 \hat F^{\dagger}_{{\bf j}+{\bf y}+{\bf x},\lambda_v,e_{v}} +
\alpha_1 \hat F^{\dagger}_{{\bf j}+{\bf y}-{\bf x},\lambda_v,e_{v}} +
\alpha_0 \hat F^{\dagger}_{{\bf j}+2{\bf y},\lambda_v,e_{v}},
\label{eq15}
\end{eqnarray} 
where $M=L/2$, ${\bf j}={\bf i}+(L/2)({\bf x}+{\bf y})$, and effectuating
the sums over the index $n$ in (\ref{eq15}), the periodic boundary conditions 
must be taken into account.

\begin{figure}[h]
\epsfxsize=6cm
\centerline{\epsfbox{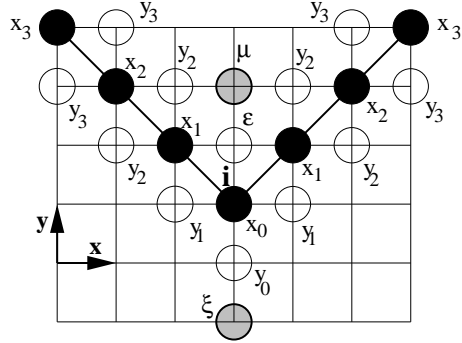}}
\caption{The first cusp region of $\hat W^{\dagger}_{v=1,{\bf i},\tau=1,
\lambda_v,e_{v}}$ constructed along the broken line (thick line) labelled by 
$({\bf i},\tau_{\bf i}=1)$. 
White, gray and full circles represent in order $\hat D^{\dagger}$,
$\hat F^{\dagger}$, and $\hat P^{\dagger}$ operators, the coefficients 
($x_l,y_l,\epsilon$, etc.) are the numerical prefactors.}
\label{fig5}
\end{figure}
\begin{figure}[h]
\epsfxsize=6cm
\centerline{\epsfbox{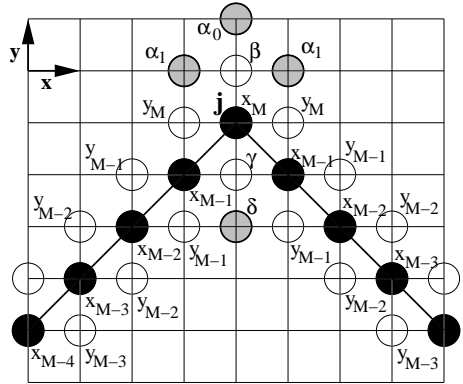}}
\caption{The second cusp region of $\hat W^{\dagger}_{v=1,{\bf i},\tau=1,
\lambda_v,e_{v}}$. The thick line is a part of the broken line. 
For notations see Fig.5.}
\label{fig6}
\end{figure}
\begin{figure}[h]
\epsfxsize=6cm
\centerline{\epsfbox{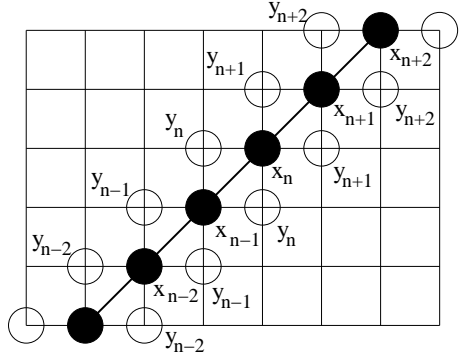}}
\caption{Inter-cusps (straight) region of 
$\hat W^{\dagger}_{v=1,{\bf i},\tau=1,\lambda_v,e_{v}}$. The thick line is 
a part of the broken line. Notations as in Fig.5.}
\label{fig7}
\end{figure}

A similar procedure based on $\tau_{\bf i}=4,\tau_{\bf j}=2$ leads
to the broken line used in constructing 
$\hat W^{\dagger}_{v,{\bf i},\tau=4,\lambda_v,e_{v}}$, which can be 
obtained from $\hat W^{\dagger}_{v,{\bf i},\tau=1,\lambda_v,e_{v}}$ by the
interchange ${\bf x}\leftrightarrow {\bf y}$.

\subsection{The numerical prefactors of the $\hat W^{\dagger}$ operators}

The numerical prefactors present in Figs.5-7 and Eqs.(\ref{eq15}) are 
given as follows
\begin{eqnarray}
&&x^d_{M \geq n \geq 0} = (-1)^{n+1}(n+1)y_0 \frac{a_{1,d}}{a_{2,d}}, 
\: y_{M \geq n \geq 1} = (-1)^n y_0,
\nonumber\\
&&x^f_0= \frac{y_0}{a_{1,f}a_{2,d}} (a^2_{1,d} + 4 a^2_{2,d}), \:
x^f_1= \frac{2y_0}{a_{1,f}a_{2,d}} (a^2_{2,d}-a^2_{1,d}), 
\nonumber\\
&&x^f_{M-2 \geq n\geq 2} = - \frac{a_{1,d}}{a_{1,f}} x^d_n, \: 
\epsilon = -3 y_0,
\: \mu = -\xi = \frac{a_{2,d}}{a_{1,f}} y_0,
\nonumber\\
&&\alpha_1 = - \frac{a_{2,d}}{a_{1,f}} (y_M - \frac{a_{2,d}}{a_{1,d}} x^d_M),
\quad \beta= -\frac{a_{2,d}}{a_{1,d}} x^d_M , 
\nonumber\\
&&\gamma= - \frac{a_{2,d}}{a_{1,d}} (2 x^d_{M-1} + x^d_M), \quad
\delta = - \frac{a_{2,d}}{a_{1,f}} (\gamma + 2 y_{M-1}), 
\nonumber\\
&&x^f_{M-1}= - \frac{a_{1,d}}{a_{1,f}} x^d_{M-1} - \frac{a_{2,d}}{a_{1,f}}
(2 y_{M-1}+y_M +\gamma),
\label{eq16}
\\
&&x^f_{M}= - \frac{a_{1,d}}{a_{1,f}} x^d_{M} - \frac{a_{2,d}}{a_{1,f}}
(2 y_{M}+\beta +\gamma), \: \alpha_0 =  \frac{a^2_{2,d} x^d_M}{a_{1,d} 
a_{1,f}},
\nonumber
\end{eqnarray}
where $y_0$ is arbitrary, and $x_i=(x^d_i,x^f_i)$. 

The deduction procedure of (\ref{eq16}) is the following one: 1) One fixes a 
$\hat W^{\dagger}_{v,{\bf i},1,\lambda_v,e_{v}}$ operator labelled by the 
fixed site ${\bf i}$ of the sublattice $v$ as shown by Figs.5-7. 2) After this 
step one takes all operators $\hat A_{{\bf i'},v',e'}$ of the Hamiltonian 
from Eq.(\ref{eq6}) which share common sites with $\hat W^{\dagger}_{
v,{\bf i},1,\lambda_v,e_{v}}$, 3) calculate the anti-commutators 
$\{ \: \hat A_{{\bf i'},v',e'} \: , \: 
\hat W^{\dagger}_{v,{\bf i},1,\lambda_v,e_{v}} \: \}$
, and 4) require zero value for these, e.g. 
\begin{eqnarray}
\{ \: \hat A_{{\bf i'},v',e'} \: , \: 
\hat W^{\dagger}_{v,{\bf i},1,\lambda_v,e_{v}} \: \}
\: = \: 0,
\label{eq17}
\end{eqnarray}
for all possible values of all indices. 
These conditions provide a linear system of equations, whose linearly 
independent terms build up the following system of equations:

In the region of the lower cusp one has
\begin{eqnarray}
&&a_{1,f} \xi + a_{2,d} y_0 = 0,
\nonumber\\
&&\epsilon a_{2,d} + a_{1,d} x^d_0 + a_{1,f} x^f_0 = a_{2,d} y_0,
\nonumber\\
&&a_{1,f} \mu + a_{2,d} \epsilon + 2 y_0 a_{2,d} =0,
\nonumber\\
&&a_{2,d} (y_0+\epsilon) + a_{1,d} x^d_1 + a_{1,f} x^f_1 =0,
\nonumber\\
&&a_{2,d} x^d_0 + a_{1,d} y_0 = 0,
\nonumber\\
&&a_{1,d}x^d_2 + a_{1,f} x^f_2 = 0,
\nonumber\\
&&a_{1,d}x^d_3 + a_{1,f} x^f_3 = 0,
\nonumber\\
&&\epsilon a_{1,d} + a_{2,d} (x^d_0 + 2 x^d_1) = 0,
\nonumber\\
&&a_{2,d} (x^d_0+x^d_1) + a_{1,d} y_1 =0, \quad y_1 = - y_0,
\nonumber\\
&&a_{2,d} (x^d_1+x^d_2) + a_{1,d} y_2 =0, \quad y_2 = y_0,
\nonumber\\
&&a_{2,d} (x^d_2+x^d_3) + a_{1,d} y_3 =0, \quad y_3 = - y_0,
\nonumber\\
&&a_{2,d} (x^d_3+x^d_4) + a_{1,d} y_4 =0, \quad y_4 = y_0.
\label{eq18}
\end{eqnarray}

For the inter-cusp region presented in Fig.7 one obtains for $4 \leq m 
\leq (M-2)$
\begin{eqnarray}
&&a_{1,d} x^d_m + a_{1,f} x^f_m = 0,
\nonumber\\
&&a_{2,d} (x^d_m + x^d_{m+1}) = - a_{1,d} y_{m+1},
\nonumber\\
&&y_m =(-1)^m y_0 .
\label{eq19}
\end{eqnarray}

The upper cusp region provides the equations
\begin{eqnarray}
&&a_{1,f} \delta + a_{2,d} \gamma = - 2 a_{2,d} y_{M-1},
\nonumber\\
&&2 a_{2,d} y_M + a_{2,d} (\beta+\gamma)+a_{1,d} x^d_M + a_{1,f} x^f_M =0,
\nonumber\\
&&a_{1,f} \alpha_0 + a_{2,d} \beta = 0,
\nonumber\\
&&a_{2,d} (\gamma + y_M) + a_{1,f} x^f_{M-1} = - 2 a_{2,d} y_{M-1} - a_{1,d}
x^d_{M-1},
\nonumber\\
&&a_{2,d} (\beta + y_M) = - a_{1,f} \alpha_1,
\nonumber\\
&&a_{2,d} ( y_M + y_{M-1}) =0,
\nonumber\\
&&a_{2,d} x^d_M + a_{1,d} \gamma = -2 a_{2,d} x^d_{M-1},
\nonumber\\
&&a_{2,d} x^d_M + a_{1,d} \beta = 0,
\label{eq20}
\\
&&a_{2,d} x^d_M + a_{1,d} y_M = - a_{2,d} x^d_{M-1}  
\nonumber
\end{eqnarray} 
The system of equations Eqs.(\ref{eq18},\ref{eq19},\ref{eq20}) must be solved 
considering $y_0$ a given, but arbitrary quantity. The solution is presented 
in Eq.(\ref{eq16}).

\subsection{The properties of the $W^{\dagger}$ operators}

The properties related to the constructed $W^{\dagger}_{v,{\bf i},
\tau,\lambda_v,e_v}$ operators are the following ones:

a) Since $L^2/2$ different possible sites ${\bf i}\in S_v$ are present in a 
given sublattice $v$, the operator 
$\hat \phi^{\lambda_v}_{e_v}(v)  =
\prod_{{\bf i}\in S_v}(\hat W^{\dagger}_{v,{\bf i},
\tau=1,\lambda_v,e_v} \hat W^{\dagger}_{v,{\bf i},\tau=4,\lambda_v,e_v})$ 
introduces $L^2$ electrons into the system. Hence, based on  $[ \prod_{v=1,2}
\hat \phi^{\lambda_v}_{e_v}(v)]$, the vector from (\ref{eq14}) becomes
\begin{eqnarray}
|\Psi^{\lambda_1,\lambda_2}_{e_1,e_2}\rangle=[ \prod_{v=1,2}
\hat \phi^{\lambda_v}_{e_v}(v)] |0\rangle,
\label{eq20a}
\end{eqnarray}
and has $2L^2$ electrons, e.g. is defined  at half filling.

b) The $2L^2$ operators $\hat W^{\dagger}_{v,{\bf i},\tau,\lambda_v,e_v}$ 
present in (\ref{eq20a}) are
constructed from $2L^2$ canonical Fermi operators ($\hat D_{{\bf i},
\lambda_v,e_v}$, $\hat F_{{\bf i},\lambda_v,e_v}$) \cite{canonic}, 
and are linearly 
independent, since each of them has at least one new segment \cite{Obs0} 
in its construction.

c) Since $W^{\dagger}_{v,{\bf i},\tau,\lambda_v,e_v}
W^{\dagger}_{v,{\bf i},\tau,\lambda_v,e_v}=0$ holds,
the operators $W_{v,{\bf i},\tau,\lambda_v,e_v}$ are
fermionic operators, but not satisfy canonical anti-commutation rules,
e.g. $\{ W_{v,{\bf i},\tau,\lambda_v,e_v},
W^{\dagger}_{v',{\bf i}',\tau',\lambda'_{v'},e'_{v'}} \} \ne \delta_{v,v'}
\delta_{{\bf i},{\bf i}'} \delta_{\tau,\tau'} \delta_{\lambda,\lambda'}
\delta_{e,e'}$.

d) Given by the construction of $\hat W^{\dagger}_{v,{\bf i},\tau,\lambda_v,
e_v}$ operators [e.g. (\ref{eq17})], one has 
\begin{eqnarray}
\hat G  |\Psi^{\lambda_1,\lambda_2}_{e_1,e_2}\rangle=0,
\label{eq21}
\end{eqnarray}
hence (\ref{eq14}) give the minimum energy eigenvalue (e.g. zero) of the
first term of the studied Hamiltonian from (\ref{eq6}).

e) A quite important property of the $\hat W^{\dagger}_{v,{\bf i},\tau,
\lambda_v,e_v}$ operators is that they introduce $f$ electrons only in the
sublattice $v$. This can be seen for example from Figs.(5-7) which best 
present the
$\hat W^{\dagger}$ operators, by taking into account that $\hat F^{\dagger}$
operators (containing the $\hat f^{\dagger}$ operators) 
act only on sites  depicted in these figures by gray and full 
(black) circles. Consequently, the $\hat \phi^{\lambda_v}_{e_v}(v)$ operator 
introduces $f$
electrons only in the sublattice $v$, and more than this, because of
$\hat F^{\dagger}_{{\bf i},\lambda_v,e_v} \hat F^{\dagger}_{{\bf i},\lambda_v,
e_v} =0$, the vector $\hat \phi^{\lambda_v}_{e_v}(v)|0\rangle$ will 
contain only single $f$
occupancy in the sublattice $v$, while the other sublattice $v'\ne v$ will
contain only $d$ electrons. Concluding this point, the contributions
created by $\hat \phi^{\lambda_v}_{e_v}(v)|0\rangle$ will have 
i) in $S_v$ empty sites, single
$d$ or $f$ occupancy, and double occupancy of the form $\hat D^{\dagger}_{
{\bf i},\lambda_v,e_v} \hat F^{\dagger}_{{\bf i},\lambda_v,e_v}$ on sites
${\bf i} \in S_v$, while ii) in the sublattice $S_{v'}$, $v'\ne v$ empty 
sites, and single $d$ occupancy can appear. 

f) Because of the property mentioned at point e), $|\Psi^{\lambda_1,
\lambda_2}_{e_1,e_2} \rangle$ from (\ref{eq20a})
will contain $\hat F^{
\dagger}_{{\bf i},\lambda_1,e_1}$ operators in the sublattice $v=1$, 
${\bf i}\in S_1$; $\hat F^{\dagger}_{{\bf j},\lambda_2,e_2}$ operators in the 
sublattice $v=2$, ${\bf j}\in S_2$, hence double $f$ occupancy is not present,
consequently one has
\begin{eqnarray}
U \hat U_f |\Psi^{\lambda_1,\lambda_2}_{e_1,e_2} \rangle= 0 .
\label{eq22}
\end{eqnarray}  
In conclusion, (\ref{eq14}) gives the minimum energy eigenvalue (e.g. zero) 
also for the second term of the studied Hamiltonian from (\ref{eq6}). Note that
even if $\lambda_1,\lambda_2$ are arbitrary but fixed in (\ref{eq6})
hence also in (\ref{eq21},\ref{eq22}), the parameters $e_1,e_2$ can be 
arbitrary chosen. 

g) As seen in Figs.4-7 the $\hat W^{\dagger}$ operators are extended. The 
extended nature has several aspects that has to be separately stressed. 
1) First, it is not possible to find $2L^2$ solutions of (\ref{eq17}) for 
$\hat W^{\dagger}$ operators such to obtain $\hat W^{\dagger}$ 
enclosed in a cluster 
smaller than the zise of the system, and to preserve the property
(\ref{eq22}). 2) By plotting all $\hat W^{\dagger}$ 
operators present for a fixed $L$ one realizes that for two arbitrary sites 
$({\bf j}_1,{\bf j}_2)$ of the system
there is at least one $\hat W^{\dagger}$ operator containing both sites, hence
representing a direct path between ${\bf j}_1$ and ${\bf j}_2$.

\section{The ground state at $U > 0$}

\subsection{The ground state at half filling}
 
The results presented in Sec.IV.E.  [see (\ref{eq21},\ref{eq22})] show that 
the ground state at half filling of $\hat H$ is provided by the 
$|\Psi^{\lambda_1,\lambda_2}_{e_1,e_2} \rangle$ vectors. 
Since as specified in Sec.IV.E, 
$e_1,e_2=\pm 1$ can be arbitrarily taken, the ground
state at half filling becomes
\begin{eqnarray}
|\Psi_g(2L^2)\rangle = |\Psi^{\lambda_1,\lambda_2} \rangle =
\sum_{e_1,e_2=\pm 1}|\Psi^{\lambda_1,
\lambda_2}_{e_1,e_2} \rangle,
\label{eq23}
\end{eqnarray}    
the ground state energy being $E_g(2L^2)=-2L^2 K_d$. 
Using (\ref{eq8},\ref{eq9}) one finds $K_d=(t_2/2)[(V_0/V_1)^2 +4]$, hence
$E_g(2L^2)=E_g^0(2L^2)$ holds for arbitrary $U > 0$. I stress here that since
double $f$-occupancy is not present in the interacting ground state, 
only the presence of $U>0$ is required for $|\Psi_g(2L^2)\rangle$,
hence (\ref{eq23}) is the ground state for arbitrary $U > 0$.

Defining $\hat \phi^{\lambda_v}=\sum_{e_v} \hat \phi^{\lambda_v}_{e_v}$,
the vector $\hat \phi^{\lambda_v}|0\rangle$ represents a singlet state for 
arbitrary $\lambda_v$. Since the ground state (\ref{eq23}) becomes now
$|\Psi_g(2L^2)\rangle=\prod_v \hat \phi^{\lambda_v}|0\rangle$, it
represents a singlet, non-magnetic state.
Given by the arbitrary nature of the $\lambda_1,\lambda_2$ coefficients,
one could write the ground state (\ref{eq23}) also in the form
$|\Psi_g(2L^2)\rangle = \sum_{\lambda_1,\lambda_2} c_{\lambda_1,\lambda_2}
|\Psi^{\lambda_1,\lambda_2} \rangle$ where $c_{\lambda_1,\lambda_2}$ are
arbitrary numerical coefficients. But this expression not emphasize a
supplementary degeneracy given by the parameters $\lambda_1,\lambda_2$, only 
shows that the spin axis for both $v=1,2$ can be arbitrarily chosen.   

\begin{figure}[h]
\epsfxsize=6cm
\centerline{\epsfbox{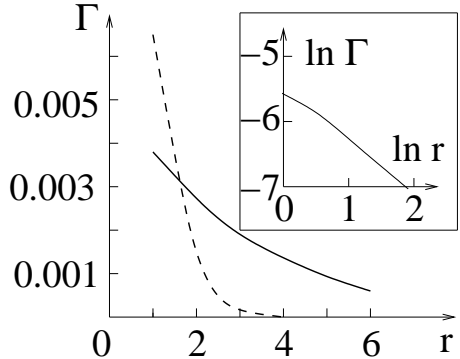}}
\caption{Ground state expectation value of f-hopping correlations
$\Gamma(\bf r)$ at $U=0$ (dashed line) and $U \ne 0$ (continuous line) for a 
$12\times 12$ system at $t_1/t_2=10,t_2/V_1=0.5,V_1/V_0=0.1$. 
${\bf r} \parallel ({\bf x}+{\bf y})$ is in $|{\bf x}+{\bf y}|$ units.
Inset shows the log-log plot at $U \ne 0$. For example,
at ${\bf r}=4$, $\Gamma_U/\Gamma_{U=0}=2 \cdot 10^5$ holds \cite{obi}.
The result is qualitatively the same for other directions and  other 
$V_1/V_0$ ratios as well.}
\label{fig8}
\end{figure}

\subsection{The one particle localization length in the interacting case}

The obtained result in (\ref{eq23}) shows that
turning on an arbitrary $U > 0$, the ground state at $\bar n=1/2$
has the property $d^f_{\bf i}=0$ on each site ${\bf i}$, where 
$d^f_{\bf i} =\langle \hat n^f_{{\bf i},\uparrow}\hat n^f_{{\bf i},\downarrow}
\rangle$, and $\langle .... \rangle$ denotes ground state expectation value
calculated in terms of $|\Psi_g(2L^2)\rangle$. Consequently, since 
$\hat d^{f}_{\bf i} > 0$ at $U=0$,
the Hubbard repulsion, by strongly 
diminishing the f-electron double occupancy, spread out 
the contributions in the ground state wave function (e.g. the
ground state in exact terms can be now given only in terms of operators
which extend along the whole system). The aspects of this fact have been
underlined in the point g) of Sec.IV.E, these properties are missing in 
the non-interacting 
case, and are introduced by the Hubbard $U > 0$. 
As a consequence, if one fixes
two arbitrary sites $({\bf j}_n,{\bf j}_m)$ of the system, there is at least 
one $\hat W^{\dagger}_{v,{\bf i}, \tau,\lambda_v,e_v}$ operator which contains
both sites, hence represents a direct path between ${\bf j}_n$, and 
${\bf j}_m$, e.g. gives
$\langle \hat W_{v,{\bf i}, \tau,\lambda_v,e_v}| \hat b^{\dagger}_{{\bf j}_n,
\sigma} \hat b_{{\bf j}_m,\sigma}|\hat W_{v,{\bf i}, \tau,\lambda_v,e_v}
\rangle / \langle \hat W_{v,{\bf i}, \tau,\lambda_v,e_v}|\hat W_{v,{\bf i}, 
\tau,\lambda_v,e_v}\rangle \sim O(1/L)$, where $|\hat W_{v,{\bf i}, \tau,
\lambda_v,e_v}\rangle = \hat W^{\dagger}_{v,{\bf i}, \tau,\lambda_v,e_v} 
|0\rangle$, and $r=|{\bf j}_n -{\bf j}_m|$ is arbitrary
\cite{obsxx}.

This property produces a clear increase
in the one particle localization length $\ell$.
I show this aspect by presenting an exemplification for $L=12$
in Fig.8, where $\Gamma({\bf r})$ from (\ref{Intr1}) calculated in terms of 
$|\Psi_g(2L^2)\rangle$ is shown.
This emphasizes that $\Gamma({\bf r})$ has no 
more strong exponential decay as in the $U=0$ non-interacting case, and
can increase even 5 
order of magnitudes given by the presence of a non-zero $U$. The $\ell$ value
corresponding to $\Gamma({\bf r})$ for the interacting case in Fig.8
(see inset) exceeds the sample size,
while at $U=0$ had the value $\ell/|{\bf x}+
{\bf y}| \sim 0.5$. The delocalization effect produced by the Hubbard $U>0$
is clearly seen.

The reason for this behavior is that $U$ throw the electrons about in order
to diminish considerably the double occupancy. This is why $\hat W^{\dagger}$
becomes extended along the whole system. In other words, it is not possible to
construct at half filling in the presence of $U>0$  $\hat W^{\dagger}$ 
operators with the properties presented in Sec.IV.E., enclosed in a cluster 
with fixed extension smaller than the sample size. 
Concerning the needed degeneracy, I note the following:
In order to construct the resulting extended
$\hat W^{\dagger}$ operator from (\ref{eq15}), the system must combine a huge 
number of linearly independent contributions. If all these contributions are 
belong to the same energy (e.g. the starting point is strongly degenerate),
the creation of the interacting ground state containing the $\hat W^{\dagger}$
operators is without cost of energy, hence it is easier to be effectuated.
This is the reason why macroscopic degeneracy is important for the process. 

\subsection{The compensation effect in the PAM}

I add at the end an observation regarding the compensation effect in the PAM
which is also visible from the obtained solution.
In the PAM, the $f$-electron moments are compensated in the singlet ground 
state at $\bar n =1/2$. Starting from the paper of Nozieres \cite{obsxy}, 
the global or local nature of this compensation effect, especially in the 
large $U$ limit, is strongly debated. The presented results show that for 
arbitrary large repulsive $U$ values the compensation can be manifestly global,
being given by a subtle superposition effect of arbitrary spin orientations,
involving both $b=d,f$ electrons.

\section{Summary and Conclusions}

In the study of the delocalization effect of the Hubbard interaction $U>0$, 
two aspects have to be analyzed in order to understand the physical reasons of 
this process: i) how $U$ extends the one particle states, and ii) how, given
by $U$, the gap in the charge excitation spectrum disappears. For the point
ii) the results deduced for the ionic Hubbard model at half filling
\cite{garg,dago,scal} suggest that the correlation screening of the one-body
potential responsible for the insulating phase can be invoked as an intuitive 
possibility for explanation. This aspect however not emphasizes clearly the
physical reasons for point i), namely, how the Hubbard
repulsion extends the one-particle behavior in a genuine multielectronic and 
interacting case.

The presented paper provide physical insights regarding this aspect i). First
one shows that always, when the delocalization effect of the Hubbard repulsion
is observed, $U$ acts on the background of a macroscopic degeneracy in a 
multiband system. After this step the delocalization effect is analyzed
produced on a band insulator starting point taken for the non-interacting case.
For this one considers a prototype two band model at half
filling, which at $U=0$ has a completely filled lowest band, separated by a gap
from the empty upper band. In order to have the macroscopic degeneracy 
present, the lowest band has been chosen flat. After this step $U$ has been
turned on, and its effect analyzed. The extended nature of the
one-particle behavior was tested by the ground state expectation value of 
the long-range hopping terms.   

This analyzis is made in the presented paper based on
the periodic Anderson model (PAM) in two dimensions and half filling 
as a prototype model. In order to provide valuable results, the study is 
presented at exact level. The PAM has 
been chosen since several aspects are known regarding the construction of its
ground states \cite{intro4,intro5}, but, exact ground states for it in 2D,
interacting case, and half filling, are deduced here for the first time.
This is achieved by the use of extended operators, the technique being 
described in the context of non-integrable models, again for the first time 
in this paper.  

The deduced results show that acting on the background of a macroscopic 
degeneracy, the Hubbard repulsion, by diminishing the double occupancy,
spread out the contributions in the ground state wave function. 
By this, is able to increase considerably 
the one-particle localization length, hence the
one-particle behavior becomes extended, which represents a clear physical 
source of the delocalization effect.

The study of the point ii) mentioned previously, and related to the gap
disappearance needs the deduction of the low laying excitation spectrum 
in the presented frame. The description of this case, at the level of the 
same quality, remains a challenging subject for future investigations.

\acknowledgements

I am indebted to D. Vollhardt, A. P. Kampf, A. Zawadowski, and M. Gulacsi 
for support and discussions. Financial support has been provided by 
Alexander von Humboldt foundation, and contract OTKA-T48782, 76821 of 
Hungarian Scientific Research Fund.


\begin{references}
\bibitem{canonic}
The total number of these operators is $4L^2$.
\bibitem{obi}
For mathematical simplicity, $\lambda_2=-1/\lambda_1^{*}$ (e.g.
opposite spins in $v=1,2$) has been considered
during the numerical calculation based on (\ref{eq23}) of Fig.8. 
Taking other $(\lambda_1,\lambda_2)$, the plot remain qualitatively the same.
\bibitem{obsxx}
Note that $O(4L)$ operators $\hat B^{\dagger}$, $B=D,F$ are present in a
$\hat W^{\dagger}$ operator, and all coefficients in (\ref{eq16}) have
roughly the same order of magnitude fixed by  $y_0$.
\end{references}
\end{document}